# Walking with the atoms in a chemical bond : A perspective using quantum phase transition

Sabre Kais*
Department of Chemistry
Purdue Quantum Science and Engineering Institute,
Purdue University,
West Lafayette, IN-47907, USA

*Classical phase transitions, like solid-liquid-gas or order-disorder spin magnetic phases, are all driven by thermal energy fluctuations by varying the temperature. On the other hand, quantum phase transitions happen at absolute zero temperature with quantum fluctuations causing the ground state energy to show abrupt changes as one varies the system parameters like electron density, pressure, disorder, or external magnetic field. Phase transitions happen at critical values of the controlling parameters, such as the critical temperature in classical phase transitions, and system critical parameters in the quantum case. However, true criticality happens only at the thermodynamic limit, when the number of particles goes to infinity with constant density. To perform the calculations for the critical parameters, finite size scaling approach was developed to extrapolate information from a finite system to the thermodynamic limit. With the advancement in the experimental and theoretical work in the field of ultra-cold systems, particularly trapping and controlling single atomic and molecular systems, one can ask: do finite systems exhibit quantum phase transition? To address this question, finite size scaling for finite system was developed to calculate the quantum critical parameters. Recent observation of a quantum phase transition in a single trapped $^{171}Yb^+$ ion in the Paul trap indicates the possibility of quantum phase transition in finite systems. This perspective focuses on examining chemical processes at ultracold temperature as quantum phase transitions -- particularly the formation and dissociation of chemical bonds – which is the basic process for understanding the whole of chemistry.*

Quantum phase transitions happen at zero temperature with quantum fluctuations. As a consequence of Heisenberg's uncertainty principle, the position and the velocity of a quantum object cannot both be measured precisely, causing the ground state energy to show abrupt changes as one varies the system parameters [1]. For example, one can look at the melting of Wigner lattice (see Figure 1), as suggested theoretically by the physicist Eugene Wigner, a metal that typically conducts electricity could become an insulator when the density of electrons is reduced at ultracold temperature. What recently has been observed experimentally by Park and Demler [2] is the melting of the crystal state into liquid because of quantum fluctuations near absolute zero temperature. These results highlight the beautiful fundamental fact of wave-particle duality in quantum mechanics, wherein electrons behave as "particle-like" in the solid phase, but as "wave-like" in the melted liquid phase. In quantum spin systems, continuous and discontinuous quantum phase transitions have been studied as a function of varying external magnetic fields, pressure, and disorder [3]. In analogy with the critical point at the classical liquid-gas phase transition in water, Jimenez et al. [4] provided experimental evidence of a quantum critical point in the geometrically frustrated quantum antiferromagnet $SrCu_2(BO_3)_2$ by controlling both the pressure and the applied magnetic field at low temperature. These phase transitions are associated with singularities of the free energy that occur only in the thermodynamic limit, where the system size approaches infinity [5]. Fisher and others [6,7] developed the finite-size scaling, a systematic theoretical approach that allows one to extrapolate information on the criticality of the finite system to the thermodynamic limit [8].

However, theoretical and experimental results indicate the existence of quantum phase transitions of finite systems. Plenio and coworkers [9] show that the Quantum Rabi Model (QRM), a two-level atom interacting with a single-mode cavity field, exhibits a continuous quantum phase transition as one varies a control parameter g (proportional to the ratio of the coupling constant to the square root of the product of the atomic transition frequency and the cavity frequency [9]). They prove that in the limit where the atomic transition frequency in the unit of the cavity frequency tends to infinity, the system has two phases: the normal phase $g < g_c = 1$ and the superradiant phase $g > g_c = 1$ (see Figure 1). Recently, Cai et al. [10] experimentally observed this continuous quantum phase transition in a system composed of only the two-hyperfine atomic levels interacting with a single-mode bosonic field using the $^{171}Yb^+$ ion in a Paul trap. They measured the spin-up populations in the two hyperfine states in the ground state manifold $^2S_{1/2}$ and the rescaled photon number associated with the spatial motion of the ion along one of its principal axes. They found that the motional degree of freedom can be well described as a quantum harmonic oscillator and thus serve as the Bosonic mode in the quantum Rabi model. This exciting new experimental observation of quantum phase transition in finite systems opens up a new field of research understanding in quantum critical phenomena of finite systems at ultra-cold temperature.

Theoretically, quantum critical phenomena and symmetry breaking of electronic structure configurations of finite systems have been studied extensively for decades. Herschbach and coworkers [11] pioneered the field of dimensional scaling for electronic structure calculations, and have discussed the symmetry breaking of electronic structure configurations of single atoms and simple molecular systems [11,12]. This approach generalizes the Schrodinger equation to the large-dimensional space and solves the resulting simple equation at the limit as D-> ∞. At this limit, the electrons take a fixed position in the large-D effective potential, allowing one to study the criticality and symmetry breaking as one varies the parameters controlling the system, such as the nuclear charge, interatomic distance in diatomic molecules, and external electric and magnetic fields [11,13]. For example, symmetry breaking at the large-D limit of electronic structure configurations of atoms and simple molecular systems has one-to-one mapping to the criticality of mean-field theory, in analogy to the mean-field criticality of fluid and magnetic systems [13]. To carry out the calculations to the physical space, D=3, we have shown that one can describe such quantum phase transitions of finite systems in Hilbert space with the analogy between the thermodynamic limit and the size of the Hilbert space [14]. To obtain the exact quantum critical parameters, a finite-size scaling can be formulated in the Hilbert space by replacing the number of particles with the number of basis functions used to obtain the exact ground-state wave function (see Figure 2). This approach can be used to describe symmetry breaking of electronic structure configurations and quantum criticality of atomic and molecular systems as one varies parameters in the Hamiltonian [15].

Next, we would like to argue that at ultracold temperatures, one can describe the formation/dissociation of chemical bonds as a quantum phase transition between a free atomic state phase and a molecularly bonded state phase. By opening a new way of examining the formation/dissociation of chemical bonds as quantum phase transitions, one might envision a universal classification of chemical systems. To illustrate this approach, we examine the bond formation of a simple generic homonuclear dimer $A_2$ formed from two atomic constituents of A. In the bonded configuration, the total ground state electronic energy of the system $A_2$ without correction due to vibrational degrees of freedom can be computed from either a simple Hartree-Fock treatment under Born-Oppenheimer approximation and refined subsequently with advanced post-Hartree Fock methods to include static and dynamic correlation. Just like in the QRM model, one can construct a fictitious two-level system in this problem wherein one of the 'levels' corresponds to the aforesaid bound state energy of $A_2$ at the equilibrium bond length and the other being the ground state energy computed similarly for the free atoms of A. The energy separation between the two forms the excitation energy in the two-level system just like in QRM. The interaction of this two-level system with the vibrational stretch mode of $A_2$ can now be envisioned to include Non-Born Oppenheimer effects. The said stretch mode with its characteristic frequency under harmonic approximation will form the bosonic reservoir in QRM and the coupling parameter g in QRM will be replaced by electronic-vibrational coupling strength. After mapping the details of the problem to the QRM model one can then show through a simple calculation that the critical point $g_c=1$ corresponds to the equilibrium distance (R=$R_{eq}$) in the formation of the molecular state of $A_2$ (see Figure 2). So far, we did not discuss the electron and nuclear spin interaction in the case of the formation of $A_2$, but one can add the spin degrees of freedom to the analysis and examine the hyperfine splitting changes from free atoms to the formation of molecules. Adding the spin and the rotational degrees of freedom [17] might give an exotic quantum molecular phase. The analysis of the chemical bond formation and breaking as a quantum phase transition at absolute zero temperature is general for any two-level model system interacting with an environment.This we believe would shed unforeseen insight into the very process of forming a chemical bond which is at the heart of chemistry and molecular physics. Further experimentation is required to investigate the idea using optically controlled atomic beams we shall discuss in the next paragraph. One can similarly describe the formation of chemical bonds in any dimeric system, resonance in benzene molecule, two base states for the molecule of the dye magenta, nitrogen tunneling in ammonia, to name just a few systems.

For many decades, the scientific community has witnessed many Nobel prizes being granted for the basic inventions and discoveries in the area of low-temperature physics, the development of methods to cool and trap atoms with laser light, Bose-Einstein condensation in dilute gases of alkali atoms, and recently for groundbreaking inventions in the field of laser physics and application of optical tweezers. The field of cold chemistry is an exciting field of research [18,19], particularly for quantum information and computing science where quantum mechanical wavelike behavior plays a central role, such as superposition, inference, and entanglement [20,21]. With the advancement of this field, recently Liu et al. [22] presented an experimental study where they combined exactly two atoms in a single, controlled reaction. The experimental apparatus traps two individual laser-cooled atoms, sodium and one cesium, in separate optical tweezers and then merges them into one optical dipole trap forming one molecule. Thus, methods of trapping and controlling single atoms to form molecules might open many exciting avenues for research, particularly quantum phase transitions and the formation of chemical bonds, and the possibility of creating exotic molecular quantum phases.

In summary, the very process of chemical bond formation , cornerstone to the foundation of chemistry , can be viewed using the perspective of a quantum phase transition.  An idea which thereby unifies the two disciplines and paves the road to further research to consolidate this newly emergent intuition.

- **Email:** kais@purdue.edu
  https://www.chem.purdue.edu/kais/

**Acknowledgment**

I would like to thank Dr. Manas Sajjan for reading and commenting on the draft. We acknowledge funding by the U.S. Department of Energy (Office of Basic Energy Sciences) under Award No. DE-SC0019215, the National Science Foundation under Award No. 1955907, NSF grant **#** 2124511, CCI Phase I: NSF Center for Quantum Dynamics on Modular Quantum Devices (CQD-MQD) and the U.S. Department of Energy, Office of Science, National Quantum Information Science Research Centers, Quantum Science Center.

**Figure Captions**

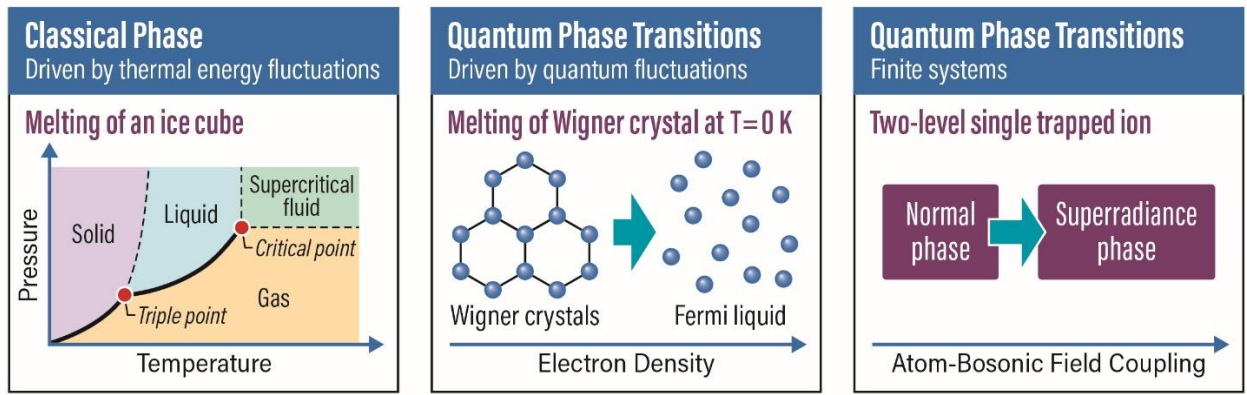

**Figure (1)** (Left) Classical phase transition in a macroscopic system like water as characterized in the (T,P) plane. The regimes wherein liquid water, solid ice and water vapor is stable are represented in three different colors whereas the black curves indicate phase boundaries wherein two phases can co-exist in equilibrium. (Middle) Melting of Wigner crystal to a Fermi liquid due to increasing electron density. Being a quantum phase transition, such transformations can occur even at absolute zero due to quantum fluctuations. (Right) Quantum phase transition (QPT) in a finite system like a single two-level system coupled with an external Bosonic reservoir. Unlike in classical phase transitions wherein macroscopic variables like pressure, temperature etc are usually involved , for the control variables for a QPT like this is simply the energy separation of the system relative to that of the bath.

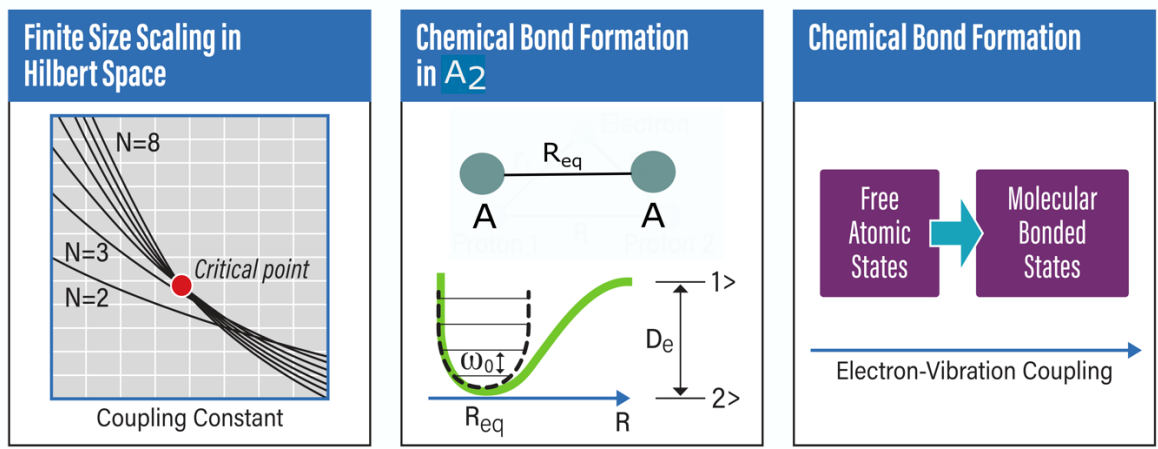

**Figure (2)** (Left) Detection of critical parameters in a Quantum Phase transition from Finite-Size Scaling approach. This can be reliably done without approaching the limit of infinitely many degrees of freedom but by successive enhancement of the dimension of the Hilbert space. (Middle) The geometric depiction of a generic $A_2$ dimer and a single phonon reservoir to which it can be coupled. (Right) The two-level description of the chemical bond formation process in $A_2$ wherein the one of the accompanying states is the free (A,A) units and the other is the bonded system.